\documentclass[10pt, conference, letterpaper]{IEEEtran}
\IEEEoverridecommandlockouts
\usepackage{cite}

\usepackage{amsmath,amssymb,amsfonts}
\usepackage{amsthm}
\usepackage{bm}
\usepackage{booktabs}
\usepackage{algorithm}
\usepackage[noend]{algpseudocode}
\usepackage{graphicx}
\usepackage{textcomp}
\usepackage{xcolor}
\usepackage{indentfirst} 
\usepackage{breqn}
\usepackage{subcaption}
\usepackage{pslatex}
\usepackage{enumitem}
\usepackage{booktabs}
\usepackage{siunitx}
\usepackage{threeparttable}
\def\BibTeX{{\rm B\kern-.05em{\sc i\kern-.025em b}\kern-.08em
    T\kern-.1667em\lower.7ex\hbox{E}\kern-.125emX}}
\begin{document}

\title{Analyzing the Impact of Adversarial Attacks on C-V2X–Enabled Road Safety: An Age of Information Perspective\\
{}
}

\author{\IEEEauthorblockN{Mahmudul Hassan Ashik}
\IEEEauthorblockA{\textit{Cyber Security Engineering Department} \\
\textit{George Mason University}\\
Fairfax, USA \\
mashik@gmu.edu}
\and
\IEEEauthorblockN{Moinul Hossain}
\IEEEauthorblockA{\textit{Cyber Security Engineering Department} \\
\textit{George Mason University}\\
Fairfax, USA \\
mhossa5@gmu.edu}
}

\maketitle

\begin{abstract}
The Cellular Vehicle-to-Everything (C-V2X), introduced and developed by the 3GPP, is a promising technology for the Autonomous Driving System (ADS). C-V2X aims to fulfill the Service-Level Requirements (SLRs) of ADS to ensure road safety following the development of the latest version, i.e., the NR-V2X. However, vulnerabilities threatening road safety in NR-V2X persist that have yet to be investigated. Existing research primarily evaluates road safety based on successful packet receptions. In this work, we propose a novel resource starvation attack that exploits vulnerabilities in the resource allocation of NR-V2X to diminish the required SLRs, making the road condition unsafe for autonomous driving. Furthermore, we establish the Age of Information (AoI) as the predominant metric for estimating the impact of adversarial attacks on NR-V2X by constructing a Discrete-time Markov chain (DTMC) based analytical model and validating it through extensive simulations. Finally, our analysis underscores how the proposed attack on NR-V2X can lead to unsafe driving conditions by reducing the SLR of time-sensitive applications in ADS up to $15\%$ from the target. Additionally, we observe that even benign vehicles act selfishly when resources are scarce, leading to further safety compromises.   

\end{abstract}

\vspace{-0.1in}

\section{Introduction} \vspace{-0.05in}
With advancements in communication and sensing technologies, V2X communication is a key enabler of the evolving, safer, and more reliable Intelligent Transport System (ITS)\cite{r1}. Both IEEE and 3GPP have pioneered standards for V2X communication: the 802.11-based Dedicated Short Range Communication (DSRC) and C-V2X, respectively. 3GPP first introduced C-V2X services in Release 14 based on Long Term Evolution (LTE) \cite{r4}. However, to further enhance LTE-V2X capabilities, 3GPP proposed NR-V2X \cite{r7}. 

\textbf{Motivation:} Although NR-V2X introduces major improvements over LTE-V2X, it still harbors exploitable vulnerabilities \cite{ying2024literature,bitsikas2025security}. An intelligent adversary can manipulate the Medium Access (MAC) layer resource allocation to degrade the Quality of Service (QoS) of ADS applications. Unlike conventional Denial-of-Service (DoS) attacks, such adversarial actions do not fully disrupt communication but instead induce controlled delays sufficient to degrade the measured timely delivery, causing it to fall below the target SLRs \cite{5GAA2023_CV2X_UseCases_Vol1}, thereby jeopardizing vehicular safety. To safeguard safety-critical ADS operations against such adversarial attacks, it is essential to investigate this emerging class of MAC-layer threats, particularly those that exploit the resource allocation mechanism.

Moreover, existing performance metrics fail to identify these threats. For instance, Packet Reception Ratio (PRR) fails to account for the lack of resources in the MAC layer, and Channel Busy Ratio (CBR), constrained by local sensing and limited observation range, is inherently unreliable for detecting such attacks \cite{yoon2020balancing}. Thus, an adversary aiming to exploit MAC layer vulnerabilities, e.g., during resource allocation, can remain undetected, as such actions do not significantly alter PRR, CBR, or related performance metrics. Therefore, such conventional metrics are inadequate for assessing the impacts of adversarial attacks on the MAC layer. In contrast, AoI \cite{kaul2012real} has emerged as a new metric for quantifying information freshness, defined as the time elapsed since the generation of the most recently received packet \cite{r10}. By capturing timeliness rather than merely reception rate or channel occupancy, AoI is inherently sensitive to subtle scheduling manipulations and resource starvation that leave PRR and CBR seemingly intact. Hence, an intelligent adversary that selectively exhausts resources will manifest as elevated AoI, making it an effective performance metric for detecting and quantifying such threats in safety-critical ADS environments. This motivates us to propose AoI as a metric for analyzing the impact of adversarial attacks that lead to unsafe road conditions.

\textbf{Challenges:} Designing intelligent attacks on NR-V2X MAC layer resource allocation poses particular challenges. Firstly, the adversary must be equipped with standard V2X equipment and mimic legitimate vehicles. This includes the transmission of legitimate Sidelink Control Information (SCI) messages at an allowable periodicity, while honoring other resource reservations at the same priority level, and so on, to ensure maximum stealthiness. Secondly, the latest 3GPP releases have already addressed vulnerabilities, such as the 1000 ms sensing window, which was exploited by \cite{twardokus2023toward} to perform a resource starvation attack. Thirdly, the adversary must be able to probe the system to determine the resource exhaustion rate, i.e., the amount of available resources that must be made unavailable to achieve the target attack success, such as creating an unsafe road condition.

\textbf{Contributions:} To fill the existing research gaps, we investigate how an adversary can take advantage of the MAC layer vulnerabilities of NR-V2X and attack the resource selection procedure, creating a resource unavailability scenario. Furthermore, we demonstrate that the adversary can significantly increase AoI by selecting different resource-exhaustion rates, thereby compromising the road safety of NR-V2X communication. The major contributions of this work are as follows:
\begin{itemize}
    \item We propose a novel attack strategy and demonstrate how an adversary can disrupt safety-critical applications by performing the attack, compromising road safety.
    \item We develop a DTMC-based analytical model to evaluate the impact of adversarial attacks on road safety and validate it by WiLabV2X-sim \cite{todisco2021performance} simulation.
\end{itemize} \vspace{-0.05in}

\section{RELATED WORKS} \label{rel_work} \vspace{-0.05in}
\textbf{Adversarial attacks on C-V2X:} Adversarial behavior in C-V2X and LTE-V2X has been examined in several recent works. The authors of~\cite{twardokus2023toward} analyzed vulnerabilities in LTE-V2X and demonstrated how an adversary can induce resource starvation by exploiting weaknesses in the sidelink resource allocation process. The work of~\cite{kim2024fatal} investigated DoS attacks on C-V2X, deriving conditions under which specific vehicles are most vulnerable and proposing a targeted “fatal” attack strategy that inflicts more severe degradation than conventional DoS methods. The study in~\cite{pethHo2024quantifying} further quantifies the impact of DoS attacks on C-V2X using PRR and end-to-end latency as primary performance indicators. Moreover, \cite{emran2025tpm} presents a remote attestation framework designed to secure 5G core network functions. Furthermore, \cite{ashik2024phantom,ashik2025reaperpulse} investigates adversarial attacks on 5G and proposes defense techniques using Markov Decision Process (MDP).

While these studies provide important insights, their evaluations predominantly rely on PRR, end-to-end latency, and update delay. Update delay, in particular, reflects how quickly packets are delivered but does not directly characterize the freshness of the received information, which is crucial for assessing the severity of the attack on C-V2X applications.

\textbf{AoI as a metric:} Several works have investigated C-V2X performance using analytical models and various metrics. The authors of~\cite{r20} developed an analytical model to evaluate the delay performance of sensing-based semi-persistent scheduling (SPS) in Mode 4. They optimized protocol parameters and validated their findings via simulation using PRR, collision probability, and CBR as key metrics. The work in~\cite{r22} proposed an analytical LTE-V2X model based on DTMC and characterized performance primarily through delay and CBR. Similarly, the authors of~\cite{cao2023semi,molina2023impact} focused on MAC-layer behavior using PRR as the principal performance metric. The study in~\cite{r23} examined SPS in NR-V2X and provided configuration guidelines for switching between SPS and dynamic scheduling (DS) under diverse traffic conditions, while the authors of~\cite{gu2022performance} analytically modeled SPS in C-V2X to assess communication reliability and latency. The authors of~\cite{r24} evaluated NR-V2X performance in terms of per-UE average throughput as a function of transmitter–receiver distance using 5G-LENA simulations. More recently, the authors of~\cite{bezmenov2024age} derived a closed-form approximation for the stationary distribution of the average AoI under SPS, and the work in~\cite{rolich2025trade} characterized the trade-off between resource reuse efficiency and timeliness in 5G NR-V2X by comparing SPS and DS, using PRR and the probability of AoI threshold violations to capture performance. 

However, despite these advances, existing studies either treat AoI as a theoretical performance indicator or rely on traditional metrics such as PRR, delay, CBR, and throughput, and none establish an AoI-driven, safety-centric framework capable of detecting intelligent MAC-layer attacks that remain undetected with conventional metrics alone.

\vspace{-0.075in}
\section{Background}
\label{sec:background}\vspace{-0.05in}

\subsection{SPS in C-V2X Mode 2}
\label{subsec:sps} \vspace{-0.05in}

C-V2X Mode~2 operation relies on SPS over the PC5 interface. Most of the physical and MAC-layer parameters and resource allocation schemes for C-V2X mode 2 (NR-V2X) are similar to those for mode 4 (LTE-V2X). However, However, NR-V2X provides additional features. In the autonomous resource allocation of  NR-V2X, a UE selects time–frequency resources from an RRC-configured \emph{sidelink resource pool} and then reuses them semi-persistently to support periodic BSM messages. The resource pool specifies numerology, subchannels, control/data resources (PSCCH/PSSCH), etc., along with a permitted list of reservation intervals that dictate how often a UE may repeat transmissions. When suitable resources become available, the UE performs sensing-based selection whenever possible, i.e., maintaining a flexible recent sensing history to rank candidate resources and avoiding those that have been observed as reserved. SCI conveys information by announcing the scheduled resources so neighbors can infer occupancy. Semi-persistence depends on the probability of reusing the same resource and is triggered when the re-selection counter (RC) reaches zero, decreasing after each transmission \cite{r9}. \vspace{-0.05in}

\subsection{Age of Information (AoI)}
\label{subsec:aoi-basics}\vspace{-0.05in}

\begin{figure}[!t]
\centering
\centerline{\includegraphics[width=0.7\columnwidth]{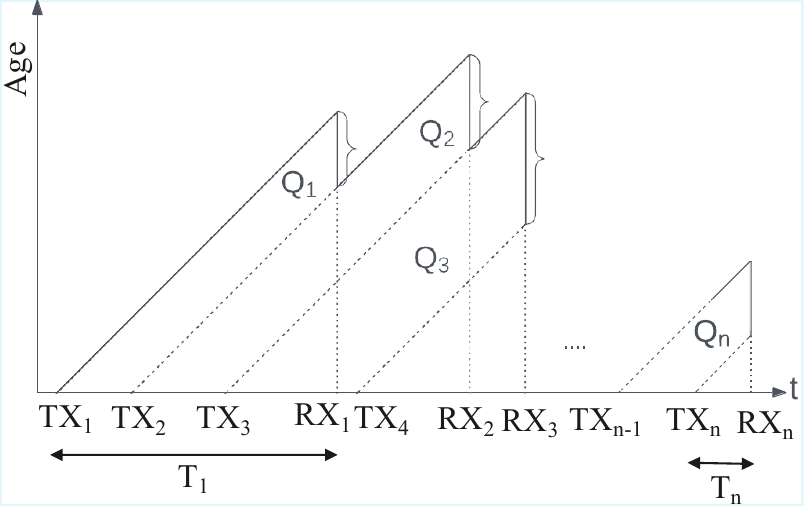}}\vspace{-0.05in}
\caption{Age of Information under a sawtooth evolution.}\vspace{-0.25in}
\label{fig_aoi}
\end{figure}

AoI measures how \emph{fresh} the receiver’s knowledge is. At time \(t\), the AoI equals the time elapsed since the most recently \emph{received} update was \emph{generated}. Fig.~\ref{fig_aoi} shows that between receptions, AoI drops when a new update is decoded correctly.

Let packet \(i\) be generated at \(Tx_i\), and let packet \(i{-}1\) be received at \(Rx_{i-1}\). Immediately after the reception of packet \(i\), the AoI resets to the inter-update gap 

\begin{equation}
\Delta(Rx_i) \;=\; Rx_i - Tx_{i-1}.
\label{eq_aoi}
\end{equation}
\vspace{-0.075in}

Thus, once we characterize (i) the time between successful receptions and (ii) the reset level at each reception, the time-average AoI follows from standard renewal expressions.

\vspace{-0.05in}
\section{Proposed Threat Model}
\label{sec:system-threat}\vspace{-0.05in}

\subsection{System Model}
\label{subsec:system} \vspace{-0.025in}

We consider C-V2X Mode~2, which utilizes the SPS mechanism. Time is slotted into subframes of duration \(\delta=1\)~ms. Each subframe offers \(M\) orthogonal \emph{subchannels} in frequency. A BSM occupies one subframe in time and exactly \(K\) \emph{contiguous} subchannels in frequency. There are \(N\) legitimate UEs (vehicles), indexed by \(u\in\{1,\dots,N\}\), each generating saturated periodic BSM traffic (i.e., a packet is always ready). An RRI of duration \(T_{\text{RRI}}\) contains
\[
\Gamma \;\triangleq\; \frac{T_{\text{RRI}}}{\delta}
\]
\vspace{-0.05in}

\noindent subframes (e.g., \(T_{\text{RRI}}{=}100\)~ms implies \(\Gamma{=}100\)). We index one RRI by a countdown \(\tau\in\{\Gamma{-}1,\ldots,0\}\).

\paragraph*{\textbf{Traffic Model}} Let \( P_{sch} \in [0,1]\) denote the probability that an allocation attempt after one sensing window finds a suitable Candidate Subframe Resource (CSR) for transmission. On a successful (re)allocation, the UE chooses a wait state randomly from \(\{0,\ldots,\Gamma-1\}\) uniformly distributed wait states. After that, the UE draws a reselection counter \(RC\) uniformly from the integers in \([R_{\min}, R_{\max}]\) and decrements \(RC\) by one after each transmission. When \(RC=0\), the UE makes a keep-or-release decision at the reservation boundary: it \emph{keeps} the current CSR with probability $P_{keep}$ (immediately reinitializing \(RC\) with a fresh draw and continuing on the same CSR every \(T_{\text{RRI}}\) ms), or it \emph{releases} the CSR with probability \(1-P_{keep}\), enters an \emph{Idle/sensing} phase, and performs sensing over a full \(\Gamma\)-subframe window to select a new CSR before resuming periodic transmissions. We assume saturated periodic packet generation: whenever a UE holds a CSR at $\tau{=}0$, exactly one packet is available. Therefore, AoI calculation is driven by (i) whether a CSR is held at $\tau{=}0$ and (ii) how long the UE remains in Idle before a successful allocation of CSR. \vspace{-0.05in}

\subsection{Proposed Threat Model}
\label{subsec:threat} \vspace{-0.05in}

As the adversary, we consider Eve, a protocol-aware attacker whose goal is to utilize the vulnerabilities of the SPS mechanism. We assume Eve is equipped with general V2V capabilities and complies with the C-V2X standard specifications, such as using the standard transmission power, reselection counter, and standard priority level, to appear legitimate. However, Eve must exploit the flexibility to set a low periodicity for BSM message transmission, as proposed in the latest release of C-V2X mode 2 operation. Let's explain how Eve can exploit this vulnerability:

\paragraph*{\textbf{Periodic Resource Starvation}} According to the latest SPS mechanism, vehicles put more weight on the observations made in the last 100 ms, instead of 1000 ms, to determine the suitable candidate CSRs. However, the latest SPS mechanism also allows a vehicle to choose the Resource Reselection Interval, i.e., the transmission periodicity as low as $1$ ms \cite{r30}. For instance, Eve can choose a very low RRI value and transmit several times using different resources each time with a standard priority level. A benign vehicle performing sensing at that time will see these resources used by Eve as “reserved" and will discount them as suitable resources. Thus, by tuning the transmission periodicity and reserving different CSRs in each transmission, Eve can render a portion of the resources unavailable, forcing the legitimate vehicles to compete for a very limited pool of available resources. Assuming Eve removes a long-run fraction $x\in[0,1]$ of usable resources by transmitting at various intervals, thereby lowering the chance that the reselection of a benign vehicle finds a free candidate CSR. Thus, in the presence of Eve, the independent per-ms Bernoulli trials with success $P'_{\text{sch}}$, where \vspace{-0.1in}

\begin{align}
    P'_{\text{sch}} \;=\; (1-x)\,P_{\text{sch}} \label{Psch_adv}
\end{align}\vspace{-0.1in}

\noindent denotes the probability of finding a suitable CSR after accounting for the attack's effects. Here, an important distinction from a jammer is that Eve does \emph{not} corrupt packets already scheduled on held CSRs; she only reduces availability at reselection, interfering with the AoI, which we formulate next. \vspace{-0.15in}

\section{Proposed Analytical Model} \vspace{-0.05in}
\label{sec:analysis}
In this section, we formulate our proposed analytical model, i.e., AoI as a metric to express the effects of the adversarial attack. Moreover, we utilize the DTMC model to explain the SPS mechanism under adversarial attack. \vspace{-0.05in}

\subsection{DTMC Model} \vspace{-0.05in}
\label{subsec:dtmc}
\begin{figure}[t] 
\centering
\includegraphics[width=0.9\columnwidth]{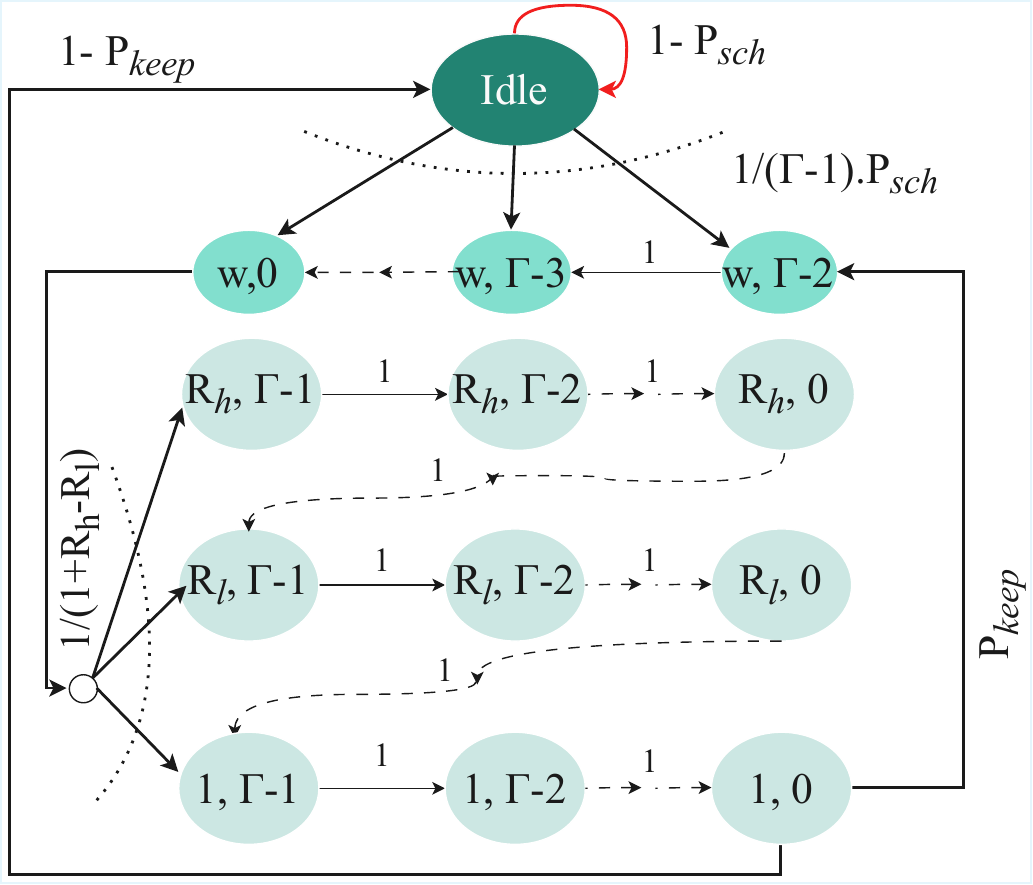} 
\caption{Proposed DTMC of the SPS mechanism.}\vspace{-0.15in}
\label{fig:dtmc-schematic}
\end{figure} 

The goal of the DTMC model is to derive the average AoI, which is defined as the expected latency from packet generation (immediately after a reselection decision) to the first allowable transmission, i.e., a transmission where the packet can secure a suitable CSR at $\tau{=}0$. As illustrated in Figure \ref{fig:dtmc-schematic}, the DTMC has (i) \emph{RRI states} $(k,\tau)$ with $\tau\in\{\Gamma{-}1,\ldots,0\}$ and $k$ the remaining $\mathrm{RC}$; and (ii) an \emph{Idle} state. Transitions $(k,\tau)\!\to\!(k,\tau{-}1)$ implement the countdown; at $(k,0)$: if $k{>}1$ then $(k,0)\to(k{-}1,\Gamma{-}1)$; else (RC expiry) keep with prob.\ $P_{\text{keep}}$ or release to Idle with prob.\ $1-P_{\text{keep}}$. In Idle, independent per-ms attempts succeed with probability $P'_{\text{sch}}$. Upon success, the UE aligns (uniform $W$) and then must traverse a full RRI to reach the next $\tau{=}0$. Additionally, each transmission attempt succeeds independently with probability \(\phi\in(0,1]\), which captures PHY decoding, interference, etc. \vspace{-0.05in}

\subsection{Average AoI Calculation} \vspace{-0.05in}

\subsubsection{Inter-success Statistics}
\label{subsubsec:inter-success}
Let \(H\) be the number of \(\tau{=}0\) attempts up to the next success: \vspace{-0.05in}
\[
H \sim \mathrm{Geom}(\phi),\qquad \mathbb{P}[H=h]=(1-\phi)^{h-1}\phi,\; h=1,2,\ldots
\]
\vspace{-0.05in}

\noindent Because attempts occur once per RRI, the inter-success time: 

\vspace{-0.05in}
\begin{equation} 
T \;=\; \Gamma\,H,
\qquad
\mathbb{E}[T] \;=\; \frac{\Gamma}{\phi},
\qquad
\mathbb{E}[T^2] \;=\; \Gamma^2\,\frac{2-\phi}{\phi^2}.
\label{eq:inter-success}
\end{equation}

\subsubsection{AoI of Successfully Received Packets}
\label{subsubsec:aoi_successful}
We model the AoI \emph{right after} a successful reception as: \vspace{-0.05in}

\begin{align}
    \mathbb{E}[D] \;=\; C_0\big(P'_{\mathrm{sch}}\big) \;+\; \big(\mathbb{E}[H]-1\big)\,\Gamma \\
\;=\; C_0\big(P'_{\mathrm{sch}}\big) \;+\; \Big(\frac{1}{\phi}-1\Big)\Gamma\;
\label{eq:reset-D}
\end{align}
\vspace{-0.05in}

We traverse the DTMC model for the benign scenario from Figure \ref{fig:dtmc-schematic} to calculate $C_0\big(P'_{\mathrm{sch}}\big)$. As explained in section \ref{subsec:system}, assuming saturated traffic, from the Idle state, with $P'_{sch} = P_{sch} = 1$, the packet secures a CSR. Then, it goes to any one of $(\Gamma-1)$ wait states with uniform probability $\frac{1}{\Gamma-1}$. After traversing all states up to $(w,0)$, the resource reservation for the packet is broadcast. Assuming the transition from the Idle state occurs after spending $1$ ms there, the average time elapsed till now:
\begin{align}
    \dfrac{1}{P'_{sch}} + \dfrac{\Gamma-1}{2} = 1 + \dfrac{\Gamma-1}{2}.
\end{align}

After that, the SPS mechanism randomly selects an RC counter value, and a total of $100$ states (from $(\Gamma-1)$ to $0$) are traversed before finally transmitting the packet using the reserved resource. Therefore, the average time elapsed: \vspace{-0.05in}
\begin{align}
    \dfrac{1}{P'_{sch}} + \dfrac{\Gamma-1}{2} + \Gamma.
\end{align}

Additionally, under the saturated traffic assumption, the maximum queue wait time is $1$ ms. Thus, for the benign scenario, assuming the packet is successfully received and decoded at the receiver, the average AoI of the successfully received packets of the benign scenario is denoted by $C_0\big(P_{\mathrm{sch}}\big)$: 
\begin{align}
    C_0\big(P'_{\mathrm{sch}}\big) = \dfrac{1}{P'_{sch}} + \dfrac{\Gamma-1}{2} + \Gamma + 1. \label{eq_aoi_dtmc}
\end{align}
\vspace{-0.1in}

\subsubsection{Formulation of AoI}
\label{subsubsec:aoi_formulation}
For a sawtooth AoI that grows linearly and resets to \(D\) at each success, renewal theory gives \vspace{-0.05in}

\begin{equation}
\bar{\Delta}
\;=\; \mathbb{E}[D] \;+\; \frac{\mathbb{E}[T^2]}{2\,\mathbb{E}[T]}.
\label{eq:renewal-aoi}
\end{equation} \vspace{-0.075in}

\noindent Substituting \eqref{eq:inter-success} and \eqref{eq:reset-D} yields

\begin{align}
\bar{\Delta}
&=  C_0\big(P'_{\mathrm{sch}}\big) \;+\; \Big(\tfrac{1}{\phi}-1\Big)\Gamma
\;+\; \Gamma\Big(\tfrac{1}{\phi}-\tfrac{1}{2}\Big),
\label{eq:aoi-expanded}\\[-2pt]
&= \;  C_0\big('P_{\mathrm{sch}}\big) \;+\; \Gamma\!\left(\frac{2}{\phi}-\frac{3}{2}\right) \; .
\label{eq:aoi-final}
\end{align} \vspace{-0.075in}

\noindent In benign scenario, from \eqref{eq:reset-D} and \eqref{eq:aoi-final}
\[
\mathbb{E}[D]\approx  C_0\big(P'_{\mathrm{sch}}\big),
\qquad
\bar{\Delta}\approx C_0\big(P'_{\mathrm{sch}}\big)+\frac{\Gamma}{2}.
\] 

On the other hand, in the attacking scenario, the availability of CSR is reduced with the probability of $P'_{sch}$ with the factor $x$ in equation \eqref{Psch_adv}. Thus, equation \eqref{eq:aoi-final} gives the average AoI calculation that not only considers the successfully received packets, but also the packets that are dropped and contribute to the average AoI calculation. Thus, we capture both benign and adversarial effects using AoI as a metric. \vspace{-0.05in}

\section{Result Analysis} \vspace{-0.05in}
\label{sec:result}
In this section, we present analytical and simulation results evaluating the AoI performance of our proposed jamming attacks. It also examines how AoI can reveal the effects of the attack on various conditions, such as the number of vehicles, and the $P_{keep}$ probability. Furthermore, we analyze how unsafe road conditions can arise from the attack on C-V2X.

\begin{table}[h] \vspace{-0.05in}
\centering
\caption{Simulation Parameters (NR--V2X Mode~2)}
\label{tab:sim-params-mode2}
\renewcommand{\arraystretch}{0.9}
\setlength{\tabcolsep}{0pt}
\begin{tabular}{@{}p{0.23\linewidth}@{\hspace{2pt}}p{0.75\linewidth}@{}}
\toprule
\textbf{Scenario} & Highway, 3 lanes; density 100 vehicles/km; avg.\ speed 70 km/h; RAW \(\in\)\{50,150,300\}~m \\
\textbf{Data} & Periodicity 100~ms; packet size 150~B (BSM) \\
\textbf{NR--V2X PHY} & SCS 15~kHz; subchannel-size 10; MCS~4; \#subchannels=5 \\
\textbf{Mode~2 MAC} & \(p_{\mathrm{keep}} = 0.4\); RC \([5,15]\); sensing threshold \(-126\)~dBm \\
\textbf{Timing} & RRI \(T_{\mathrm{RRI}} = 100\)~ms \(\Rightarrow \Gamma = 100\) subframes\\
\textbf{Adversary} & Resource starvation fraction \(x \in \{0.5,0.8,0.9\}\)  \\
\bottomrule
\end{tabular}
\end{table} \vspace{-0.2in}

\subsection{Simulation Setup}
 \label{subsec:simulation_setup}\vspace{-0.05in}
 We validate our analytical model using \texttt{WiLabV2Xsim}, an open–source, event–driven simulator that emulates PC5 sidelink with flexible numerology and realistic PHY/MAC procedures, as explained in Table \ref{tab:sim-params-mode2}. Here, we enable C-V2X Mode~2 operation and extract the per-packet AoI. \vspace{-0.05in}

\subsection{Validation of the Proposed Analytical Model} 
\begin{figure}[!t] 
  \centering
    \subfloat[]{\includegraphics[width=0.48\columnwidth]{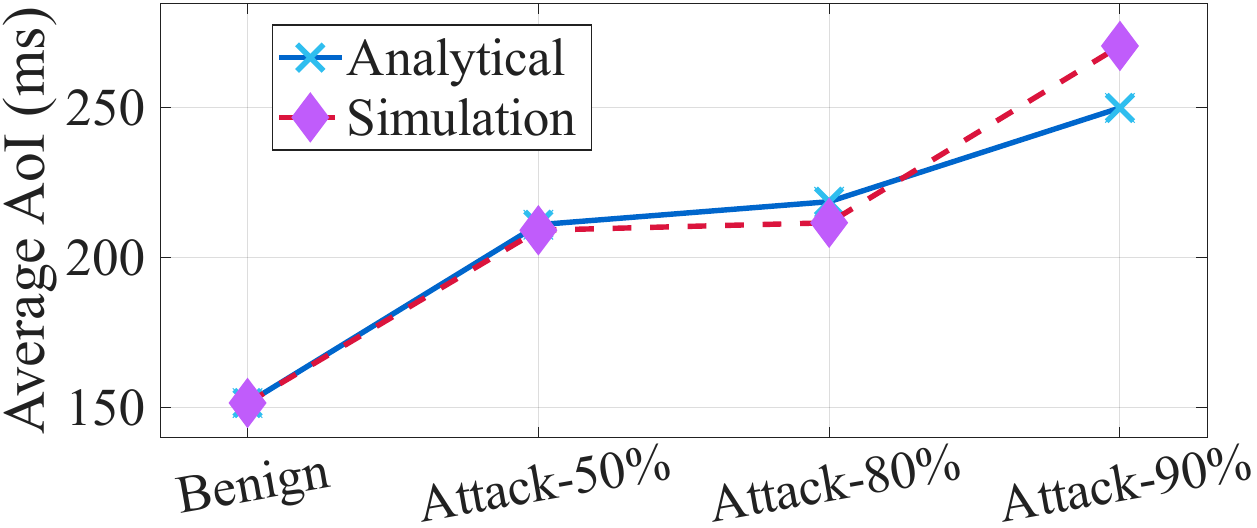}} 
  \hfill
    \subfloat[]{\includegraphics[width=0.48\linewidth]{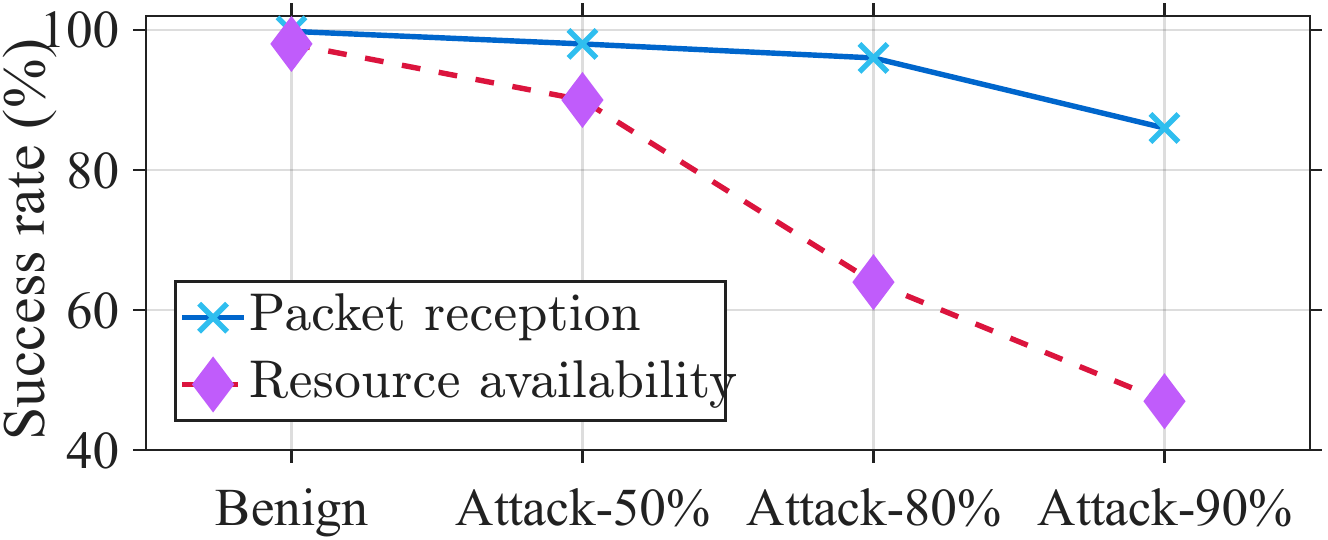}} 
\caption{(a) Validation of the proposed analytical model; (b) Limitation of PRR to show the adversarial effects.}\vspace{-0.2in}
\label{fig_val_prr}
\end{figure}

\vspace{-0.05in}
\label{subsec:val}
To validate our proposed analytical model, we compare the simulation results with those obtained from the analytical model using exact parameters. Specifically, analytical AoI is calculated from equation~\eqref{eq:aoi-final}. From Fig.~\ref{fig_val_prr}(a), we can see a close match between analysis and simulation across benign and adversarial cases \(x\in\{0,0.5,0.8,0.9\}\), where $x$ denotes the resource starvation fraction.
Overall, the gap is small (typically \(\lesssim 1\%\) in benign and \(\lesssim 6\%\) at the harshest attack). We observe that the average AoI increases convexly with the attack factor.

On the other hand, Fig~\ref{fig_val_prr}(b) shows how the packet reception success, i.e., PRR, does not correctly indicate resource exhaustion as it only considers the transmissions that managed to secure resources. However, the attacker reduces the number of available resources, as seen from the resource availability visualization of Fig~\ref{fig_val_prr}(b), which PRR cannot account for. Therefore, AoI is crucial for analyzing MAC-layer attacks.

\textbf{Insight:} Our motivation for adopting AoI as the principal performance metric stems from its ability to capture dynamics that conventional metrics overlook. As shown in Fig.~\ref{fig_val_prr}(b), Packet Reception Ratio (PRR) reflects successful receptions at the PHY layer but remains blind to variations in resource availability at the MAC layer. In contrast, AoI intrinsically links these two domains: it rises whenever a vehicle fails to access resources and falls only after a successful packet reception. This enables AoI to quantify not just communication success but also the temporal freshness of information delivery under realistic resource constraints. Therefore, unlike PRR, which merely reports success events, AoI provides a holistic view of the communication system, encompassing both MAC-layer scheduling dynamics and PHY-layer reliability—making it a more reliable and revealing metric of adversarial attacks. \vspace{-0.05in}

\subsection{AoI Variations for Different Conditions} \vspace{-0.05in}

The simulation results show that the impact of our proposed adversarial attack can vary significantly across road conditions, providing valuable insights for both attackers and defenders.

\subsubsection{AoI vs. Number of Vehicles}
\label{sec:aoi-vs-n}
\begin{figure}[!t] 
  \centering
    \subfloat[]{\includegraphics[width=.48\columnwidth]{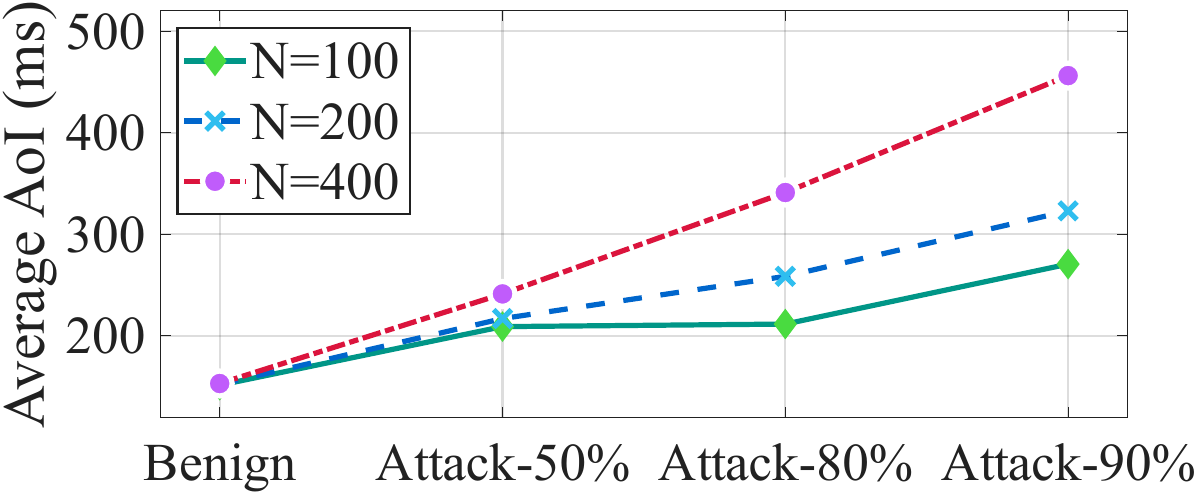}} 
    \subfloat[]{\includegraphics[width=.48\columnwidth]{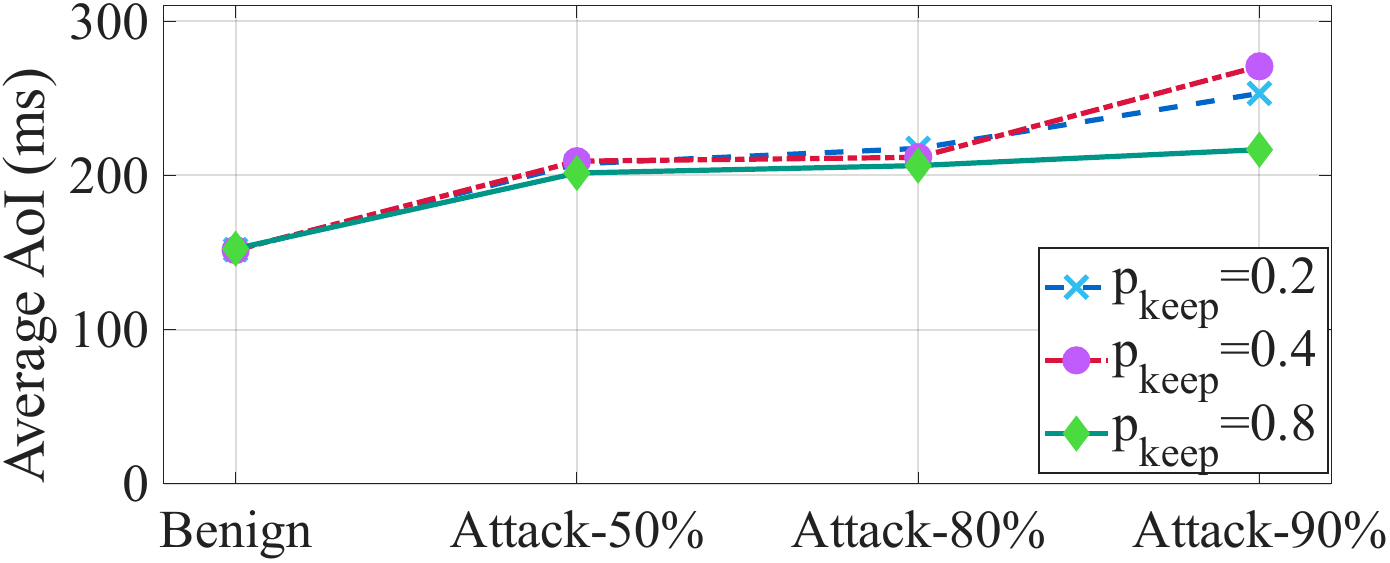}} 
\caption{(a) AoI vs. the number of vehicles; (b) AoI vs.\ \(P_{\mathrm{keep}}\).}\vspace{-0.1in}
\label{fig_aoi_n_pkeep}
\end{figure}

Fig.~\ref{fig_aoi_n_pkeep}(a) illustrates that the average AoI increases with the total number of vehicles (\(N\)) across all adversarial levels (resource exhaustion factors). This behavior reflects the compounded effects of congestion and malicious resource starvation, which increase the AoI. This inflation is most pronounced under severe attack conditions (e.g., \texttt{Attack-90\%}), where resource scarcity amplifies delay due to the inverse dependence on the available resource scheduling probability (\(1/P'_{\mathrm{sch}}\)). 
In contrast, the benign case exhibits a more gradual increase with $N$, indicating a higher likelihood of successful reselection and timely packet delivery. Therefore, increasing \(N\) decreases the probability of finding a suitable resource, thereby decreasing road safety.

\textbf{Insight:} An adversary can exploit specific environmental contexts to intensify the effect of an attack, while an effective defense strategy must dynamically adapt to those same conditions. For example, during peak traffic hours on weekdays, the increased number of vehicles leads to denser resource contention, thereby increasing the risk to road safety. Consequently, defensive mechanisms must explicitly account for these context-dependent vulnerabilities rather than rely on static countermeasures. These findings underscore the importance of environment-aware resilience strategies that can maintain road safety and communication reliability under varying traffic and attack scenarios.

\subsubsection{AoI vs. $P_{keep}$}
\label{sec:aoi-vs-pkeep}
Fig.~\ref{fig_aoi_n_pkeep}(b) presents the variation of avg. AoI with respect to the re-selection probability (\(P_{\mathrm{keep}}\)) under both benign and adversarial conditions. As \(P_{\mathrm{keep}}\) increases, vehicles tend to retain their previously selected resources for longer, leading to stable transmission opportunities and lower AoI in the benign case. However, under adversarial conditions, vulnerability arises: the attacker’s continuous occupation of resources forces legitimate nodes to re-select the same resources, causing prolonged staleness in resource reservations.

\textbf{Insight:} Even though the adversary has no control over a vehicle's $P_{keep}$ value, as a secondary effect, the proposed attack might force vehicles to choose high $P_{keep}$ values to maintain the QoS, creating a selfish resource-holding situation that can further exacerbate the impact of the attack. 

\begin{table}[!h] \vspace{-0.1in}
  \centering
  \caption{Application requirements \cite{5GAA2023_CV2X_UseCases_Vol1} vs. measured timely delivery (TDR) under \texttt{attack-90\%}.}
  \label{tab:req-vs-measured}
  \footnotesize
  \setlength{\tabcolsep}{3pt} 
  \renewcommand{\arraystretch}{1.15} 
  \begin{threeparttable}
  \resizebox{\columnwidth}{!}{%
    \begin{tabular}{l c c c c}
      \toprule
      \textbf{Service} & \textbf{AoI threshold} & \textbf{Target SLR} & \textbf{Measured TDR} & \textbf{Difference} \\
      \midrule
      Forward Collision Warning (FCW)  & 100\,ms & 99.99\%      & 84.68\% & 15.31\% \\
      Emergency Brake Warning (EBW)    & 120\,ms & 99.99\%      & 84.94\% & 15.05\% \\
      Lane Change Warning (LCW)        & 400\,ms & 99.90\%      & 86.06\% & 13.84\% \\
      \bottomrule 
    \end{tabular}%
  }
  \end{threeparttable}
\end{table}\vspace{-0.1in}

\subsection{Effects of Adversarial Attacks on C-V2X Services}
\label{subsec:effects_services}\vspace{-0.05in}
As studied in \cite{5GAA2023_CV2X_UseCases_Vol1}, there are various SLRs for C-V2X applications that require maintaining an AoI threshold beyond which service reliability degrades, leading to unsafe road conditions. For instance, Table \ref{tab:req-vs-measured} shows three services with their respective AoI threshold and target SLRs. Moreover, the AoI-violation probability, i.e., the probability that AoI exceeds the threshold, is a standard safety-critical analysis \cite{bezmenov2024age} used to analyze how unsafe road conditions may arise.
Here, we discuss how the increase in the probability of AoI violation due to the proposed attack renders the road unsafe for running advanced applications enabled by C-V2X.

\subsubsection{Road Safety Violations due to the Proposed Attack:} 
\begin{figure}[!h]
  \centering
  \subfloat[Forward Collision Warning]{%
    \includegraphics[width=.48\columnwidth]{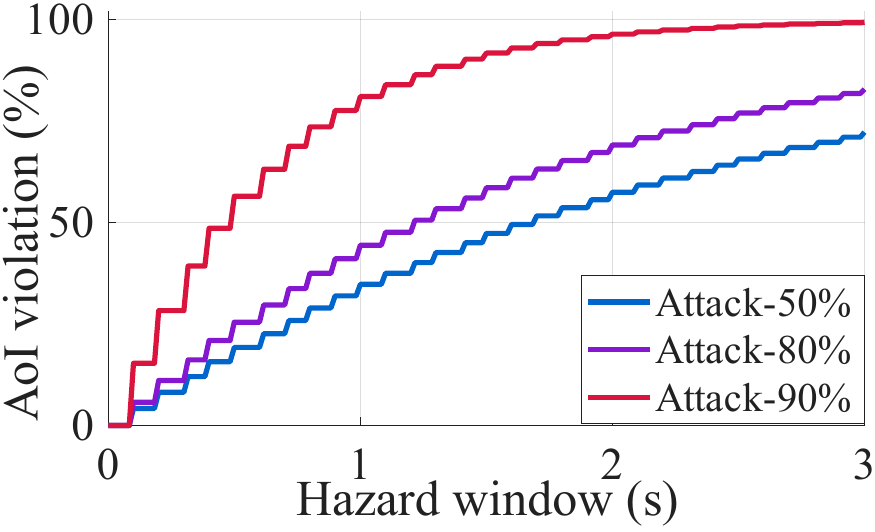}%
  }%
  \hfill
  \subfloat[Emergency Brake Warning]{%
    \includegraphics[width=.48\columnwidth]{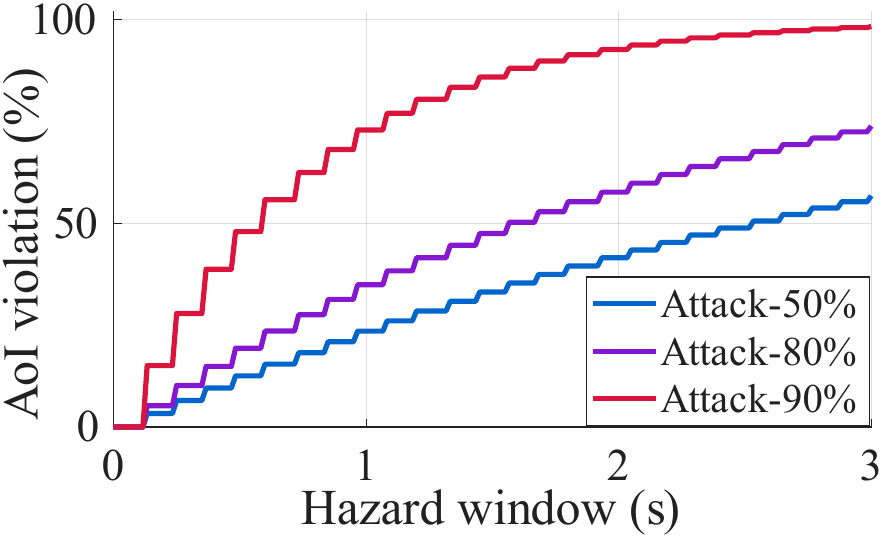}%
  } \\ 
  \subfloat[Lane Change Warning]{%
    \includegraphics[width=.48\columnwidth]{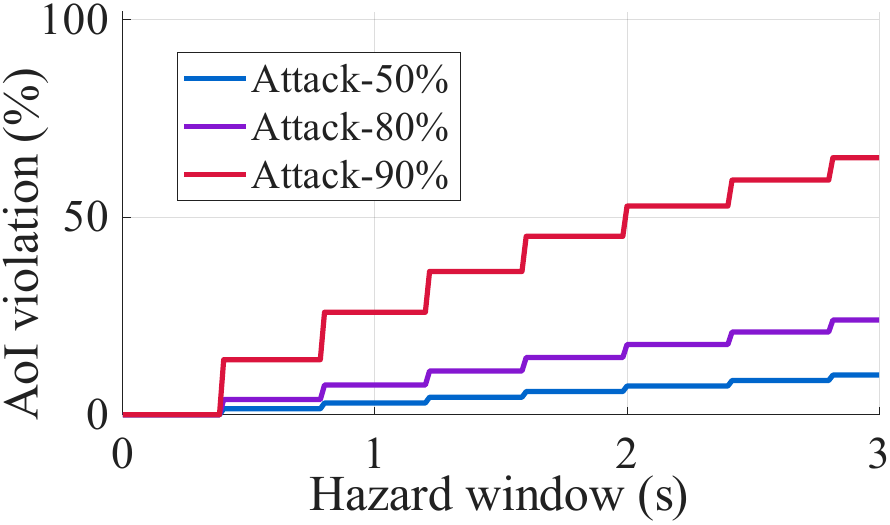}
  }%
  \subfloat[Combined]{%
    \includegraphics[width=.48\columnwidth]{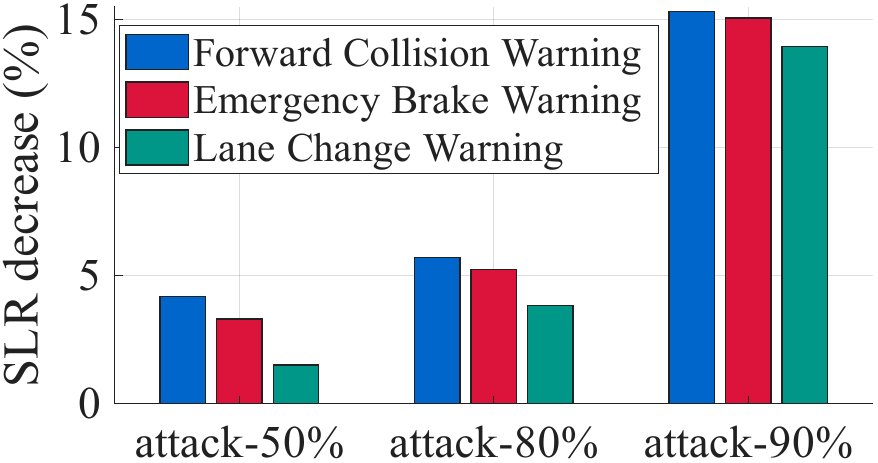}
  }%
  \caption{Probability of (a) FCW failure, 
           (b) EBW failure, 
           (c) LCW failure, and
           (d) Combined result for comparison.} \vspace{-0.15in}
  \label{fig_hazards}
\end{figure}

Fig.~\ref{fig_hazards} is constructed by mapping the measured TDR (by counting the number of packets experiencing less AoI than the threshold) in Table~\ref{tab:req-vs-measured} to the target SLR and AoI thresholds for FCW, EBW, and LCW. For each service, we use its target SLR for the adversarial scenarios (\texttt{Attack-50\%}, \texttt{Attack-80\%}, and \texttt{Attack-90\%}), to compute, over a hazard window \(h\), the probability of AoI violation. Figures~\ref{fig_hazards}(a)--(c) show this probability for FCW, EBW, and LCW, respectively. Furthermore, Fig~\ref{fig_hazards}(d) represents the decrease in reliability for different levels of the attack, showing at least nearly a $5\%$ decrease in reliability from the target SLR, rendering the road conditions for C-V2X operation significantly unsafe.

\textbf{Insight.}
Table~\ref{tab:req-vs-measured} shows that FCW requires \(99.99\%\) reliability, yet under \texttt{Attack-90\%} it drops to \(84.68\%\). Also, Fig.~\ref{fig_hazards}(a) exibits the AoI violation probability approaching \(100\%\) within a \(2\)~s hazard window. EBW exhibits the same in Fig.~\ref{fig_hazards}(b), again effectively failing to ensure road safety. Although LCW is less severely affected, Fig.~\ref{fig_hazards}(c) still shows a steep increase in the violation probability over short hazard windows, exposing vehicles to unsafe lane-change decisions. Collectively, these results confirm that C-V2X becomes significantly unsafe under the proposed resource starvation attack.

\vspace{-0.05in}
\section{Conclusion}
\label{conclusion} \vspace{-0.05in}
We developed a DTMC-based analytical framework for the SPS mechanism in NR\mbox{-}V2X to formulate the AoI under the proposed adversarial resource starvation attack and validated it against WiLabV2Xsim. Despite improvements over LTE\mbox{-}V2X, our results show that NR\mbox{-}V2X remains vulnerable at the MAC layer. Furthermore, our analysis reveals that time-sensitive services suffer an \(\approx 15\%\) reliability loss and a clear increase in AoI under the proposed attack. Moreover, we observed that AoI degrades under a resource starvation attack and improves when vehicles conserve their resources for longer periods, thereby avoiding unnecessary re-selections. This means that if an adversary can create a sufficiently persistent resource starvation, a legitimate vehicle will start acting selfishly by keeping resources for longer periods. This can lead to an even greater increase in AoI, making safety-critical autonomous systems more prone to failure. Overall, the framework exposes MAC-layer vulnerabilities in NR\mbox{-}V2X operations and highlights the need for an intelligent, proactive scheduling and detection to preserve the target service requirements for safety-critical applications in autonomous driving.\vspace{-0.05in}

\bibliographystyle{ieeetr}
\bibliography{ref}
\end{document}